\documentclass{article}

\usepackage{arxiv}

\usepackage[utf8]{inputenc} 
\usepackage[T1]{fontenc}    
\usepackage{hyperref}       
\usepackage{url}            
\usepackage{booktabs}       
\usepackage{amsfonts}       
\usepackage{nicefrac}       
\usepackage{microtype}      
\usepackage{lipsum}

\title{Empirical evaluation of full-reference image quality metrics on MDID database}

\author{
  Domonkos Varga\\
  Department of Networked Systems and Services\\
  Budapest University of Technology and Economics\\
}

\begin{document}
\maketitle

\begin{abstract}
 In this study, our goal is to give a comprehensive evaluation of 32 state-of-the-art FR-IQA metrics
 using the recently published MDID. This database contains distorted images derived from a set
 of reference, pristine images using random types and levels of distortions. Specifically, Gaussian
 noise, Gaussian blur, contrast change, JPEG noise, and JPEG2000 noise were considered.
\end{abstract}


\section{Introduction}


The goal of objective image quality assessment is to design mathematical models that
are able to predict the perceptual quality of digital images. The classification of
objective image quality assessment algorithms is based on the accessibility of the
reference image. 
In the case of reference image is unavailable image
quality assessment is considered as a no-reference (NR) one. Reduced-reference (RR)
methods have only partial information about the reference image, while full-reference
(FR) algorithms have full access to the reference image.

The research of objective image quality assessment demands databases that contain
images with the corresponding MOS values. To this end, a number of image
quality databases have been made publicly available.
Roughly speaking, these
databases can be categorized into three groups. The first one contains a smaller set
of pristine, reference digital images and artificially distorted images derived from the
pristine images considering different artificial distortions at different intensity
levels. The second group contains only digital images with authentic distortions
collected from photographers, so pristine images cannot be found in such
databases. Virtanen \textit{et al.} \cite{virtanen2014cid2013} were
first to introduce this type of database
for images by releasing CID2013. As a consequence, the development of FR methods
is connected to the first
group of databases.
In contrast Waterloo Exploration \cite{228606:5144146} and
KADIS-700k \cite{kadid10k} databases are meant to provide
an alternative evaluation of objective
image quality assessment models, by means of paired comparisons. That is why, they
contain a set of reference (pristine) images, distorted images, and distortion levels. In
contrast to other databases, they do not provide MOS values.
Information about major publicly available image quality assessment databases are
summarized in Table \ref{table:iqadatabase}.

In this study, we provide a comprehensive evaluation of 32 full-reference image quality assessment
(FR-IQA) algorithms on MDID database. In contrast to other available image quality databases,
the images in MDID contain multiple types of distortions simultaneously.

The rest of this study is organized as follows. There are a number of publicly available image quality
databases, such as IVC \cite{ivcdb}, LIVE IQA \cite{sheikh2006statistical},
A57 \cite{chandler2007vsnr}, Toyoma \cite{tourancheau2008impact},
TID2008 \cite{ponomarenko2009tid2008}, CSIQ \cite{larson2010most},
IVC-LAR \cite{ivclardb}, MMSP 3D \cite{goldmann2010impact},
IRSQ \cite{ma2012image}, \cite{ma2012study},
TID2013  \cite{ponomarenko2013color}, CID2013 \cite{virtanen2014cid2013},
LIVE In the Wild \cite{livewild},
Waterloo Exploration \cite{228606:5144146}, 
MDID \cite{sun2017mdid}, KonIQ-10k \cite{lin2018koniq},
KADID-10k \cite{lin2019kadid}, and KADIS-700k \cite{lin2019kadid}.
In Section \ref{sec:iqadatabases}, we give a brief introduction to each of them.
In Section \ref{sec:exp}, we give a comprehensive evaluation of 31 full-reference image
quality assessment
(FR-IQA) algorithms on MDID database.
Finally, a conclusion is drawn in Section \ref{sec:conc}.

\section{Image quality databases}
\label{sec:iqadatabases}

\textbf{IVC}\footnote{http://www2.irccyn.ec-nantes.fr/ivcdb/} \cite{ivcdb} 
database consists of 10 pristine images, and 235 distorted images,
including four types of distortions (JPEG, JPEG2000, locally adaptive resolution coding,
blurring). Quality score ratings (1 to 5) are provided in the form of MOS.

\textbf{LIVE Image Quality Database}\footnote{http://www.live.ece.utexas.edu/research/quality/subjective.htm} 
(LIVE IQA) \cite{sheikh2006statistical} has
two releases, Release
1 and Release 2. Laboratory for Image and Video Engineering (University of Texas
at Austin) conducted an extensive experiment to obtain scores from human subjects
for a number of images distorted with different distortion types. Release 2 has more
distortion types --- JPEG (169 images), JPEG2000 (175 images), Gaussian blur (145
images, White noise (145 images), bit errors in JPEG2000 bit stream (145 images).
The subjective quality scores in this database are DMOS (Differential MOS), ranging
from 0 to 100.

\textbf{A57 Database}\footnote{http://vision.eng.shizuoka.ac.jp/mod/page/view.php?id=26}
\cite{chandler2007vsnr}
has 3 pristine
images, and 54 distorted images, including
six types of distortions (JPEG, JPEG2000, JPEG2000 with dynamic contrast-based
quantization, quantization of the LH subbands of DWT, additive Gaussian white noise,
Gaussian blurring). Quality score ratings (0 to 1) are provided in the form of DMOS.

\textbf{Toyoma Database} \cite{tourancheau2008impact} 
consists of 14
pristine images, and 168 distorted images,
including two types of distortions (JPEG, JPEG2000). Quality score ratings (1 to 5)
are provided in the form of MOS.

\textbf{Tampere Image Database 2008}\footnote{http://www.ponomarenko.info/tid2008.htm}
(TID2008) \cite{ponomarenko2009tid2008}
contains 25 reference images
and 1,700 distorted images (25 reference images $\times17$ types of
distortions $\times4$ levels of
distortions). The MOS was obtained from the results of 838 experiments carried out
by observers from three countries. 838 observers have performed 256,428 comparisons
of visual quality of distorted images or 512,856 evaluations of relative visual quality
in image pairs. Higher value of MOS (0 - minimal, 9 - maximal, MSE of each score
is 0.019) corresponds to higher visual quality of the image. A file enclosed “mos.txt”
contains the Mean Opinion Score for each distorted image.

\textbf{Computational and Subjective Image Quality}\footnote{http://vision.eng.shizuoka.ac.jp/mod/page/view.php?id=23}
(CSIQ) \cite{larson2010most}
database consists
of 30 original images, each distorted using one of six types of distortions, each at
four to five different levels of distortion. The images were subjectively rated based on
a linear displacement of the images across four calibrated monitors placed side-by-side
with equal viewing distance to the observer. The database contains 5,000 subjective
ratings from 35 different --- both male and female --- observers. Quality score ratings
(0 to 1) are provided in the form of DMOS.

\textbf{IVC-LAR}\footnote{http://ivc.univ-nantes.fr/en/databases/LAR/} \cite{ivclardb}
database contains 8 pristine
images (4 natural images and 4 art
images), and 120 distorted images, consisting of three types of distortions (JPEG,
JPEG2000, locally adaptive resolution coding). Quality score ratings (1 to 5) are
provided in the form of MOS.

\textbf{Wireless Imaging Quality}\footnote{https://computervisiononline.com/dataset/1105138665} 
(WIQ) Database \cite{engelke2009reduced},
\cite{engelke2010subjective}  consists of 7 reference
images and 80 distorted images. The subjective quality scores are given in DMOS,
ranging from 0 to 100.

In contrast to other publicly available image quality databases \textbf{MMSP 3D Image
Quality Assessment Database}\footnote{https://mmspg.epfl.ch/downloads/3diqa/} 
\cite{goldmann2010impact} consists of stereoscopic images
with a resolution of $1,920\times1,080$ pixels.
Specifically, 10 indoor and outdoor scenes were captured
with a wide variety of colors, textures, and depth structures. Furthermore, 6 different
stimuli have been considered corresponding to different camera distances (10, 20, 30,
40, 50, and 60 cm) for each scene.

\textbf{Image Retargeting Subjective Quality}\footnote{http://ivp.ee.cuhk.edu.hk/projects/demo/retargeting/index.html}
(IRSQ) Database
\cite{ma2012image}, \cite{ma2012study}
consists of 57 reference images grouped into four attributes, specfically face and people,
clear foreground object, natural scenery, and geometric structure. Moreover,
ten different retargeting methods (cropping, seam carving,
scaling, shift-map editing, scale and
stretch, etc.) are applied to generate retargeted images.
In total, 171 test images can
be found in this database.

\textbf{Tampere Image Database 2013}\footnote{http://www.ponomarenko.info/tid2013.htm} 
(TID2013) \cite{ponomarenko2013color} contains 25
reference images
and 3,000 distorted images (25 reference images $\times24$ types of
distortions $\times5$ levels of
distortions). MOS (Mean Opinion Score) is provided as subjective score, ranging from
0 to 9.

The \textbf{CID2013}\footnote{http://www.helsinki.fi/psychology/groups/visualcognition/}
\cite{virtanen2014cid2013} database contains
474 images with authentic distortions captured by 79
imaging devices, such as mobile phones, digital still cameras, and digital single-lens reflex
cameras.

\textbf{LIVE In the Wild Image Quality Challenge Database}\footnote{http://live.ece.utexas.edu/research/ChallengeDB/} \cite{livewild}
contains
widely diverse authentic image distortions on a large number of images captured using a
representative variety of modern mobile devices. The LIVE In the Wild Image Quality
Database has over 350,000 opinion scores on 1,162 images evaluated by over 8,100
unique human observers.

\textbf{Waterloo Exploration}\footnote{https://ece.uwaterloo.ca/~k29ma/exploration/} \cite{228606:5144146}
database consists of
4,744 reference images and
94,880 distorted images created from them. Instead of collecting MOS for each test
image, the authors introduced three alternative test criteria to evaluate the performance
of IQA models, such as discriminability test (D-test), listwise ranking consistency test
(L-test), and pairwise preference consistency test (P-test).

In contrast to other databases considering artificial distortions,
\textbf{MDID}\footnote{https://www.sz.tsinghua.edu.cn/labs/vipl/mdid.html} \cite{sun2017mdid}
obtains distorted images from reference images with random types and levels of distortions.
In this way, each distorted image contains multiple types of distortions simultaneously.
Gaussian noise, Gaussian blur, contrast change, JPEG noise, and JPEG2000 noise were considered.

The main challenge in applying state-of-the-art deep learning methods to predict
image quality in-the-wild is the relatively small size of existing quality scored datasets.
The reason for the lack of larger datasets is the massive resources required in generating
diverse and publishable content. 
In \textbf{KonIQ-10k}\footnote{http://database.mmsp-kn.de/koniq-10k-database.html} \cite{lin2018koniq} a new
systematic and scalable
approach is presented to create large-scale, authentic image datasets for image quality
assessment. KonIQ-10k \cite{lin2018koniq} consists
of 10,073 images, on which large scale crowdsourcing
experiments has been carried out in order to obtain reliable quality ratings from 1,467
crowd workers (1.2 million ratings) \cite{saupe2016crowd}.
During the test users exhibiting unusual
scoring behavior were removed.

\textbf{KADID-10k}\footnote{http://database.mmsp-kn.de/kadid-10k-database.html} \cite{lin2019kadid}
consists of 81
pristine images and $10,125$ distorted images derived from the
pristine images considering $25$ different distortion types at 5 intensity levels
($10,125=81\times25\times5$). In contrast, \textbf{KADIS-700k} \cite{lin2019kadid} contains
$140,000$ pristine images and distorted images were derived using $25$ different
distortion types at 5 intensity levels but MOS values are not given in this database.

\begin{table}[h]
\caption{Major publicly available image quality assessment databases.
Publicly available image quality databases can be divided into three groups.
The first one contains a smaller set of reference images and artificially distorted images are derived from them using different noise types at
different intensity levels. There are also databases which contains only pristine images, distorted images, and distortion levels without MOS. 
} 
\centering 
\begin{center}
    \begin{tabular}{ |c|c|c|c|c|c|}
    \hline
 \footnotesize{Database} & \footnotesize{Year} & \footnotesize{Reference images} & \footnotesize{Test images} & \footnotesize{Distortion type} & \footnotesize{Subjective score}    \\
    \hline
 \footnotesize{IVC \cite{ivcdb}}                         & \footnotesize{2005} & \footnotesize{10} & \footnotesize{235}  & \footnotesize{artificial} & \footnotesize{MOS (1-5)} \\
 \footnotesize{LIVE IQA \cite{sheikh2006statistical}}                    & \footnotesize{2006} & \footnotesize{29} & \footnotesize{779}  & \footnotesize{artificial} & \footnotesize{DMOS (0-100)} \\
 \footnotesize{A57 \cite{chandler2007vsnr}}                         & \footnotesize{2007} & \footnotesize{3}  & \footnotesize{54}   & \footnotesize{artificial} & \footnotesize{DMOS (0-1)} \\
 \footnotesize{Toyoma \cite{tourancheau2008impact}}                      & \footnotesize{2008} & \footnotesize{14} & \footnotesize{168}  & \footnotesize{artificial} & \footnotesize{MOS (1-5)} \\
 \footnotesize{TID2008 \cite{ponomarenko2009tid2008}}                     & \footnotesize{2008} & \footnotesize{25} & \footnotesize{1,700}& \footnotesize{artificial} & \footnotesize{MOS (0-9)} \\
 \footnotesize{CSIQ \cite{larson2010most}}                        & \footnotesize{2009} & \footnotesize{30} & \footnotesize{866}  & \footnotesize{artificial} & \footnotesize{DMOS (0-1)} \\
 \footnotesize{IVC-LAR \cite{ivclardb}}                     & \footnotesize{2009} & \footnotesize{8}  & \footnotesize{120}  & \footnotesize{artificial} & \footnotesize{MOS (1-5)} \\
 \footnotesize{WIQ \cite{engelke2009reduced}, \cite{engelke2010subjective}}                         & \footnotesize{2009} & \footnotesize{7}  & \footnotesize{80}   & \footnotesize{artificial} & \footnotesize{DMOS (0-100)} \\
 \footnotesize{MMSP 3D \cite{goldmann2010impact}}                     & \footnotesize{2009} & \footnotesize{9}  & \footnotesize{54}   & \footnotesize{artificial} & \footnotesize{MOS (0-100)} \\
 \footnotesize{IRSQ \cite{ma2012image}, \cite{ma2012study}} & \footnotesize{2011} & \footnotesize{57} & \footnotesize{171}  & \footnotesize{artificial} & \footnotesize{MOS (0-5)} \\
 \footnotesize{TID2013 \cite{ponomarenko2013color}}                     & \footnotesize{2013} & \footnotesize{25} & \footnotesize{3,000}& \footnotesize{artificial} & \footnotesize{MOS (0-9)} \\
 \footnotesize{CID2013 \cite{virtanen2014cid2013}}                     & \footnotesize{2013} & \footnotesize{8} & \footnotesize{474}& \footnotesize{authentic} & \footnotesize{MOS (0-9)} \\
 \footnotesize{LIVE In the Wild \cite{livewild}}            & \footnotesize{2016} & \footnotesize{-}  & \footnotesize{1,162}& \footnotesize{authentic}  & \footnotesize{MOS (1-5)} \\
 \footnotesize{Waterloo Exploration \cite{228606:5144146}}        & \footnotesize{2016} & \footnotesize{4,744} & \footnotesize{94,880}&\footnotesize{artificial}& - \\
 \footnotesize{MDID \cite{sun2017mdid}}        & \footnotesize{2017} & \footnotesize{20} &\footnotesize{1600} &\footnotesize{artificial}& \footnotesize{MOS (0-8)} \\
 \footnotesize{KonIQ-10k \cite{lin2018koniq}}                   & \footnotesize{2018} & \footnotesize{-}  &\footnotesize{10,073}& \footnotesize{authentic}  & \footnotesize{MOS (1-5)}  \\
 \footnotesize{KADID-10k \cite{kadid10k}}                   & \footnotesize{2019} & \footnotesize{81} &\footnotesize{10,125}& \footnotesize{artificial}  & \footnotesize{MOS (1-5)} \\
 \footnotesize{KADIS-700k \cite{kadid10k}}                  & \footnotesize{2019} &\footnotesize{140,000}&\footnotesize{700,000}& \footnotesize{artificial}  & - \\
 \hline
 \end{tabular}
\end{center}
\label{table:iqadatabase}
\end{table}

\section{Experimental results}
\label{sec:exp}

\begin{table}
 \caption{Performance comparison of 31 FR-IQA algorithms on MDID database.}
  \centering
  \begin{tabular}{|l|l|l|l|l|}
    \toprule
    Method &Year    & PLCC     & SROCC & KROCC \\
    \midrule
BLeSS-SR-SIM \cite{temel2016bless}& 2016  &	0.7535	&0.8148	&0.6258 \\
BLeSS-FSIM	\cite{temel2016bless}& 2016 & 0.8193	&0.8467	&0.6576 \\
BLeSS-FSIMc	\cite{temel2016bless}& 2016 & 0.8527	&0.8827	&0.7018 \\
CBM	 \cite{gao2005content}&  2005       &0.7367	    &0.7212	&0.5306 \\
CSV \cite{temel2016csv}&  2016	&0.8785&	0.8814&	0.6998 \\
CW-SSIM	\cite{sampat2009complex}&2009  &0.5900	&0.6148	&0.4450 \\
DSS	\cite{balanov2015image}&2015  &0.8714	&0.8661	&0.6793 \\
ESSIM \cite{zhang2013edge}&  2013	&0.6694&	0.8253&	0.6349\\
FSIM \cite{zhang2011fsim}&  2011	&0.8591	&0.8870	&0.7074 \\
FSIMc \cite{zhang2011fsim}&  2011	&0.8639	&0.8902	&0.7122 \\
GMSD \cite{xue2013gradient}&  2013	&0.8544&	0.8617&	0.6797 \\
HaarPSI	\cite{reisenhofer2018haar}& 2018 &\textbf{0.9051}&	\textbf{0.9028}&	\textbf{0.7340}\\
MAD	\cite{larson2010most}& 2010 &0.7439	&0.7243	&0.5327\\
MCSD \cite{wang2016multiscale}&  2016	&0.8386	&0.8457	&0.6622\\
MDSI ('mult') \cite{nafchi2016mean}&  2016	&0.8130	&0.8278	&0.6441\\
MDSI ('sum') \cite{nafchi2016mean}&  2016	&0.8249	&0.8363	&0.6527\\
MS-SSIM	\cite{wang2003multiscale}&2003  &0.7884	&0.8292	&0.6360\\
MS-UNIQUE \cite{prabhushankar2017ms}& 2017 	&0.8604	&0.8712	&0.6893\\
NQM	\cite{damera2000image}& 2000 &0.6177	&0.5869	&0.4143\\
PerSIM \cite{temel2015persim}&  2015	&0.8282	&0.8196	&0.6296 \\
PSNR-HVS \cite{egiazarian2006new}	&2006  &0.679	&0.6637	&0.4845\\		
PSNR-HVS-M \cite{ponomarenko2007between}	&2007  &0.6875	&0.6739	&0.4944\\
QILV \cite{aja2006image}&  2006	&0.3296	&0.4592	&0.3214\\
QSSIM \cite{kolaman2011quaternion}&  2011	&0.8022	&0.8014	&0.6074\\
RFSIM \cite{zhang2010rfsim}& 2010 	&0.7035&	0.6758&	0.4884\\
SCIELAB	\cite{zhang1997color}& 1997 &0.2552	&0.1232	&0.0824\\
SR-SIM	\cite{zhang2012sr}& 2012 &0.7948	&0.8517	&0.6683\\
SSIM \cite{wang2004image}& 2004 	&0.5798	&0.5761	&0.4105 \\
SSIM CNN \cite{amirshahi2018reviving}	& 2018 &0.8706	&0.8804	&0.6992\\
SUMMER	\cite{temel2019perceptual}& 2019 &0.7427	&0.7343	&0.5434\\
UQI	\cite{wang2002universal}& 2002 &0.2175	&0.3608	&0.2476\\
VSI	\cite{zhang2014vsi}& 2014 &0.7883	&0.8570	&0.6710\\
    \bottomrule
  \end{tabular}
  \label{tab:table}
\end{table}

The evaluation of objective visual quality assessment is based on the correlation between
the predicted and the ground-truth quality scores. Pearson’s linear correlation
coefficient (PLCC) and Spearman’s rank order correlation coefficient (SROCC) are
widely applied to this end. Furthermore, some authors give the Kendall’s rank order
correlation coefficient as well.

The PLCC between data set $A$ and $B$ is defined as
\begin{equation}
    PLCC(A,B) = \frac{\sum_{i=1}^{n} (A_i-\overline{A})(B_i-\overline{B})}{\sqrt{\sum_{i=1}^{n}(A_i-\overline{A})^{2}}\sqrt{\sum_{i=1}^{n}(B_i-\overline{B})^{2}}},
\end{equation}
where $\overline{A}$ and $\overline{B}$ stand for the average of set $A$ and $B$, $A_i$ and $B_i$ denote the $i$th elements of set $A$ and $B$,
respectively. For two ranked
sets \textit{A} and \textit{B} SROCC is defined as
\begin{equation}
    SROCC(A,B)=\frac{\sum_{i=1}^{n} (A_i-\hat{A})(B_i-\hat{B})}{\sqrt{\sum_{i=1}^{n}(A_i-\hat{A})^{2}}\sqrt{\sum_{i=1}^{n}(B_i-\hat{B})^{2}}},
\end{equation}
where $\hat{A} $ and $\hat{B} $ are the middle ranks of set \textit{A} and \textit{B}.
KROCC between dataset $A$ and $B$ can be calculated as
\begin{equation}
 KROCC(A,B) = \frac{n_c-n_d}{\frac{1}{2}n(n-1)},
\end{equation}
where $n$ is the length of the input vectors, $n_c$ is the number of concordant pairs between $A$ and $B$, and $n_d$ is the number of discordant pairs between
$A$ and $B$.

We collected 31 FR-IQA metrics whose source codes are available online. Furthermore, we reimplemented 
SSIM CNN\footnote{https://github.com/Skythianos/Pretrained-CNNs-for-full-reference-image-quality-assessment} \cite{amirshahi2018reviving} in MATLAB R2019a.
In Table \ref{tab:table}, we present PLCC, SROCC, and KROCC values measured over the MDID database.
It can be clearly seen from the results that there is still a lot of space for the improvement of FR-IQA algorithms because only HaarPSI \cite{reisenhofer2018haar} was able to produce PLCC and SROCC values higher
than 0.9. Furthermore, only three methods --- FSIM \cite{zhang2011fsim},
FSIMc \cite{zhang2011fsim}, HaarPSI	\cite{reisenhofer2018haar} --- were able to produce KROCC values
higher than 0.7.
\section{Conclusion}
\label{sec:conc}
First, we gave information about the mostly applied
image quality databases. Subsequently, we extensively evaluated 32 state-of-the-art FR-IQA methods on MDID
database whose images contain multiple types of distortions simultaneously.
We dmonstrated that there is still a lot of space for the improvement of FR-IQA algorithms because
only HaarPSI \cite{reisenhofer2018haar} was able to produce PLCC and SROCC values higher
than 0.9.


\newpage
\bibliographystyle{unsrt}  
\bibliography{references}  






\end{document}